\documentclass[pra,aps,showpacs,amssymb,amsfonts,twocolumn]{revtex4}
\usepackage{amsmath,bbm,mathrsfs}
\usepackage{graphicx}
\usepackage{amsfonts}
\usepackage{amssymb}

\def\textbf#1{{\bf #1}}
\def\be{\begin{equation}}
\def\ee{\end{equation}}
\def\ben{\begin{eqnarray}}
\def\een{\end{eqnarray}}
\def\eea{\end{array}}
\def\bea{\begin{array}}
\newcommand{\ot}[0]{\otimes}
\newcommand{\Tr}[1]{\mathrm{Tr}#1}
\newcommand{\bei}{\begin{itemize}}
\newcommand{\eei}{\end{itemize}}
\newcommand{\ket}[1]{|#1\rangle}
\newcommand{\bra}[1]{\langle#1|}

\newcommand{\proj}[1]{\ket{#1}\bra{#1}}

\begin{document}

\title{General scheme for construction of scalar separability criteria from positive maps.}

\author{Remigiusz Augusiak}
\author{Julia Stasi\'nska}

\address{Faculty of Applied Physics and Mathematics,
Gda\'nsk University of Technology, Narutowicza 11/12, 80--952 Gda\'nsk,
Poland}

\begin{abstract}
We present a general scheme allowing for construction of scalar
separability criteria from positive but not completely positive
maps. The concept is based on a decomposition of every positive
map $\Lambda$ acting on $M_{d}(\mathbb{C})$ into a difference of
two completely positive maps $\Lambda_1$, $\Lambda_2$, i.e.,
$\Lambda=\Lambda_1-\Lambda_2$. The scheme may also be treated as a
generalization of the known entropic inequalities, which are
obtained from the reduction map. Analyses performed on a few
classes of states show that the new scalar criteria are stronger
than the entropic inequalities and when derived from
indecomposable maps allow for detection of bound entanglement.
\end{abstract}
\pacs{03.67.Mn}
\maketitle

{\it Introduction.-} Though the theory of entanglement
\cite{RevModPhys} has been developed for many years, it is still
an open problem how to unambiguously determine whether a given
quantum state is separable or not.
Recall that we call a given bipartite state separable iff it can
be written as a convex combination of product states
\cite{Werner}, i.e.,
\begin{equation}\label{separable}
\varrho_{\mathrm{sep}}=\textstyle{\sum_{i}}p_{i}\varrho_{A}^{(i)}\ot\varrho^{(i)}_{B},\qquad
\textstyle{\sum_{i}}p_{i}=1,
\end{equation}
otherwise it is called entangled. Among the known criteria
allowing for the detection of entanglement in bipartite quantum
systems, one of the most common is based on positive but not
completely positive (CP) maps \cite{maps}. It states that a given
density matrix $\varrho$ is separable if and only if the operator
$X_{\Lambda}(\varrho)=[I\ot\Lambda](\varrho)$ is positive for all
positive maps $\Lambda$ (here, $I$ denotes an identity map).
An important example of such a map is the transposition map
\cite{Peres,maps}. Among others, the so-called indecomposable
\cite{indecomp} maps are of special interest, since they can
detect bound entanglement
\cite{BE1}. However, so far only few examples of indecomposable
maps are known (see, e.g.,
\cite{Tanahasi,Osaka,Ha1,Choi,genChoi,Ha2,Kossakowski,Bcrit,Hall}).
The positive maps criterion is a structural one. To apply it one
must know the density matrix of a state.

Of interest are the criteria that can be directly applied in
experiment. One of the best known are the entropic inequalities
\cite{HHHEntr,reduction,VollbrechtWolf} saying that the entropy of
the subsystems of a separable state cannot exceed the entropy of
the full system. Mathematically it can be expressed as
\begin{equation*}
S_{\alpha}(\varrho_{A})\leq S_{\alpha}(\varrho)\quad\mathrm{and}\quad
S_{\alpha}(\varrho_{B})\leq S_{\alpha}(\varrho)\qquad (\alpha\in[0,\infty)),
\end{equation*}
where $\varrho_{A(B)}=\Tr_{B(A)}\varrho$ and as $S_{\alpha}$ one
may choose the Renyi entropy
$S_{\alpha}(\varrho)=\ln\Tr\varrho^{\alpha}/(1-\alpha)$.
Simple calculations lead to a short form of the above
inequalities, i.e.,
$\Tr\varrho_{A(B)}^{\alpha}\geq \Tr\varrho^{\alpha}$ for
$\alpha\in(1,\infty)$ and
$\Tr\varrho_{A(B)}^{\alpha}\leq \Tr\varrho^{\alpha}$ for
$\alpha\in[0,1)$. These do not constitute strong separability
criteria, for instance cannot detect bound entanglement, since
they are implied by the reduction criterion \cite{VollbrechtWolf}.
Still, their experimental realization for $\alpha=2$ is within
reach of the existing technology \cite{Bovino}.

Interesting questions arise here. {\it Is it possible to derive
inequalities such as entropic ones from any positive map not only
from the reduction one? Moreover, would such inequalities detect
entanglement more efficiently and in particular could detect bound
entanglement?} Recently, it was shown in Ref. \cite{ASH} that
imposing some conditions on a density matrix and utilizing an
extended reduction criterion \cite{Bcrit,Hall} one may derive some
entropic-like inequalities, stronger than the entropic ones. In
particular, the inequalities can detect bound entanglement.

Here we provide alternative simple construction allowing for
derivation of scalar separability criteria from any positive, not
CP map.
Surprisingly, for many positive maps the inequalities can be
derived, unlike in \cite{ASH}, for all states, without any
assumptions on commutativity.

{\it Inequalities.-} Let $\Lambda$ be a positive but not
completely positive map. A corresponding necessary separability
criterion states that if a given density matrix is separable,
i.e., of the form (\ref{separable}), then
$[I\ot\Lambda](\varrho_{\mathrm{sep}})\geq 0$.
Applying the fact that every positive but not CP map $\Lambda$
defined on $M_{d}(\mathbb{C})$ can be written as a difference of
two CP maps $\Lambda_{1}$ and $\Lambda_{2}$ (see, e.g., Refs.
\cite{Jamiolkowski, Hou}), i.e.,
\begin{equation}\label{decomposition}
\Lambda=\Lambda_{1}-\Lambda_{2},
\end{equation}
we can rewrite the separability criterion in the form of the
following operator inequality
\begin{equation}\label{op_inequality}
[I\ot\Lambda_{1}](\varrho_{\mathrm{sep}})\geq
[I\ot\Lambda_{2}](\varrho_{\mathrm{sep}}).
\end{equation}
Now, we are prepared to show the aforementioned inequalities,
which may be treated as a generalization of the standard entropic
inequalities.

{\it Theorem.} Let $\varrho$ be a separable state and $\Lambda$ a
positive but not completely positive map that can be written as in
Eq. (\ref{decomposition}). Then the following implications hold:
\begin{description}
\item[(i)] If $[[I\ot\Lambda_{2}](\varrho),\varrho]=0$ and
$\alpha\geq 0,\, \beta> 1$ then
\begin{equation}\label{th1}
\Tr\varrho^{\alpha}\left([I\ot\Lambda_{1}](\varrho)\right)^{\beta}\geq
\Tr\varrho^{\alpha}\left([I\ot\Lambda_{2}](\varrho)\right)^{\beta}.
\end{equation}
\item[(ii)] If $\alpha\geq 0$ and $0\leq \beta\leq 1$ then
\begin{equation}\label{th2}
\Tr\varrho^{\alpha}\left([I\ot\Lambda_{1}](\varrho)\right)^{\beta}\geq
\Tr\varrho^{\alpha}\left([I\ot\Lambda_{2}](\varrho)\right)^{\beta}.
\end{equation}
\item[(iii)] If $\alpha\geq 0$ and $\beta\in[-1,0)$ then
\begin{equation}\label{th3}
\Tr\varrho^{\alpha}\left([I\ot\Lambda_{1}](\varrho)\right)^{\beta}\leq
\Tr\varrho^{\alpha}\left([I\ot\Lambda_{2}](\varrho)\right)^{\beta}.
\end{equation}
\item[(iv)] If $\alpha,\beta\geq 0$ then
\begin{equation}\label{th4}
\Tr\varrho^{\alpha}\left([I\ot\Lambda_{1}](\varrho)\right)^{\beta}\geq
\Tr(\Sigma_{\downarrow}(\varrho)^{\alpha}\Sigma_{\uparrow}([I\ot\Lambda_{2}](\varrho))^\beta).
\end{equation}
\end{description}

{\it Proof.} {\bf (i)} Assuming that a given $\varrho$ is
separable, the inequality (\ref{op_inequality}) is satisfied.
Following Ref. \cite{VollbrechtWolf} we have
\begin{eqnarray}
\Tr\varrho^{\alpha}\left([I\ot\Lambda_{1}](\varrho)\right)^{\beta}&\negmedspace=\negmedspace&
\Tr\,\mathrm{e}^{\ln\varrho^{\alpha}}\mathrm{e}^{\ln\left([I\ot\Lambda_{1}](\varrho)\right)^{\beta}}\nonumber\\
&\negmedspace\geq\negmedspace&
\Tr\,\mathrm{e}^{\ln\varrho^{\alpha}+\ln\left([I\ot\Lambda_{1}](\varrho)\right)^{\beta}}\nonumber\\
&\geq& \Tr\,\mathrm{e}^{\alpha\ln\varrho+\beta\ln[I\ot\Lambda_{2}](\varrho)}.
\end{eqnarray}
Then, using the commutativity assumption we obtain
\begin{equation}
\Tr\varrho^{\alpha}\left([I\ot\Lambda_{1}](\varrho)\right)^{\beta}\geq
\Tr\varrho^{\alpha}\left([I\ot\Lambda_{2}](\varrho)\right)^{\beta},
\end{equation}
which finishes the first part of the proof.

{\bf (ii)-(iii)} It suffices to use the fact that the operator
function $f(A)=A^{r}$ is monotonically increasing for $r\in[0,1]$,
and monotonically decreasing for $r\in[-1,0)$ (cf. \cite{Bhatia}).

{\bf (iv)} We start from the fact that for any $A,B\in
M_d(\mathbb{C})$ one has
$e^{A+B}=\lim_{m\to\infty}(e^{A/2m} e^{B/m}e^{A/2m})^{m}$
(cf. \cite{Bhatia}). This, due to the continuity of the trace,
implies
\begin{eqnarray}
&&\hspace{-0.5cm}\Tr\,e^{\alpha\ln\varrho+\beta\ln[I\ot\Lambda_{2}](\varrho)}\nonumber\\
&&=\lim_{m\to\infty}
\Tr\left[\varrho^{(\alpha/2m)}([I\ot\Lambda_{2}](\varrho))^{(\beta/m)}\varrho^{(\alpha/2m)}\right]^{m}.
\end{eqnarray}
Now, we use the following inequality \cite{Lieb}:
\begin{equation}
\Tr\left[B^{r}(\sqrt{B}A\sqrt{B})^{s}\right]\geq
\Tr\left[\left(\Sigma_{\uparrow}(A))^{s}(\Sigma_{\downarrow}(B)\right)^{r+s}\right],
\end{equation}
satisfied by positive matrices $A$ and $B$, and $r,s\geq 0$. Here
$\Sigma_{\uparrow}(X)=\mathrm{diag}[\sigma_{1},\ldots,\sigma_{n}]$
and
$\Sigma_{\downarrow}(X)=\mathrm{diag}[\sigma_{n},\ldots,\sigma_{1}]$,
where $\sigma_{1}\geq \ldots\geq \sigma_{n}$ are singular values
of $X$, i.e., the eigenvalues of $|X|$. Putting $r=0$,
$A=([I\ot\Lambda_{2}](\varrho))^{\beta/m}$, and
$B=\varrho^{\alpha/m}$ we arrive at the claimed inequality. $\quad
\square$

{\it Remarks.} It should be emphasized that though parts (ii)-(iv)
of the theorem are proved without any assumptions on
commutativity, there is, to our knowledge, no direct method of
measurement of such functions (except for the case $\beta$=1 and
natural $\alpha$ included in part (ii)). Meaning that they do not
lead to many-copy entanglement witnesses, whereas the inequality
from part (i) of the theorem taken with integer $\alpha$ and
$\beta$ can be experimentally checked in a collective measurement
(see \cite{collective,ASH}). The main drawback, however, is that
to apply the criterion we need to assume that
$[[I\ot\Lambda_{2}](\varrho),\varrho]=0$. Remarkably, as we will
see in the examples given below, for many of the known positive
maps, $\Lambda_{2}$ could be taken as an identity map. Then the
commutativity assumption is trivially satisfied by every state and
the inequality (\ref{th1}) becomes
\begin{equation}\label{L2eqId}
\Tr\varrho^{\alpha}((I\ot\Lambda_{1})(\varrho))^{\beta}\geq
\Tr\varrho^{\alpha+\beta}\quad(\alpha\geq 0,\,\beta > 1).
\end{equation}

Because of their structure we shall call the inequalities from
parts (i)-(iii) of the Theorem the {\it
$(\alpha,\beta)$-inequalities}.

{\it Examples.-} Below we provide four examples of positive maps
and show that most of them have identity map as $\Lambda_{2}$ in
Eq. (\ref{decomposition}), which leads to entropic-like
inequalities (\ref{L2eqId}).

{\it Example 1.} First, we consider the reduction map
\cite{reduction,CAG} which acts on a $d\times d$ matrix $X$ as
$R(X)=(\Tr\,X)\mathbbm{1}_{d}-X$. It can be written as a
difference of CP maps $R_{1},R_{2}$ taken as
$R_{1}(X)=(\Tr\,X)\mathbbm{1}_{d}$ and $R_{2}(X)=X$. The complete
positivity of both is obvious. Moreover, $R_{2}$ is an identity
map so by the remarks given above the inequality is suitable for
all states. Putting $\alpha=1$ and using the fact that
$\Tr\varrho(\varrho_{A}\ot\mathbbm{1}_{d})^{\beta}=\Tr\varrho_{A}^{\beta+1}$,
one obtains the standard entropic inequality
$\Tr\varrho_{A}^{\beta+1}\geq \Tr\varrho^{\beta+1}$
On the other hand, as shown in Fig. 1, inequalities of the form
(\ref{L2eqId}) derived from the reduction map for $\alpha>1$ and
$\alpha >\beta$ are stronger than the entropic ones.

{\it Example 2.} The second example is concerned with the modified
transposition map $\tau^{U}$ acting on $X$ as
$\tau^{U}(X)=UT(X)U^{\dagger}$ with $T$ being the known
transposition map and $U$ denoting arbitrary $d\times d$ unitary
matrix. It may be written as a difference of
$\tau_{1}^{U}=(1/2)\tau^{U}\circ\tilde{R}$ and
$\tau_{2}^{U}=(1/2)\tau^{U}\circ R$,
where $\tilde{R}(X)=(\Tr\,X)\mathbbm{1}_{d}+ X$. Both maps
may be easily shown to be completely positive \cite{footnote}.
To check which states fulfill the assumption of the theorem we may
write
$\tau_{2}^{U}(X)=(1/2)[(\Tr\,X)\mathbbm{1}_{d}-\tau^{U}(X)]$
and then $[I\ot
\tau_{2}^{U}](\varrho)=(1/2)[\varrho_{A}\ot\mathbbm{1}_{d}-\tau_{B}^{U}(\varrho)]$,
which means that one needs to assume that $[\varrho_{A}\ot
\mathbbm{1}_{d},\varrho]=[\tau_{B}^{U}(\varrho),\varrho]$, where
$\tau_{B}^{U}=I\ot \tau^{U}$. Notice that this is the only map
studied here for which one needs to impose some assumptions on
$\varrho$ to derive inequalities (\ref{th1}) from it.

{\it Example 3.} As the third example we consider the
indecomposable map introduced in Refs. \cite{Bcrit,Hall}, that is
$\Lambda^{U}(X)=(\Tr\,X)\mathbbm{1}_{d}-\tau^{U}(X)-X$ and its
slight modification. Here $\tau^{U}$ is defined as in Example 2
but with $U$ being an antisymmetric matrix $(U^{T}=-U)$ such that $U^{\dagger}U\leq
\mathbbm{1}_{d}$. This map may be expressed as a difference of the maps
$\Lambda_{1}^{U}(X)=(\Tr\,X)\mathbbm{1}_{d}-\tau^{U}(X)$ and
$\Lambda_{2}^{U}(X)=X$. Note that the map $\Lambda_{1}^{U}$ is
the same as $\tau_2^{U}$ (up to one-half factor) and
thus is completely positive. Note that combining maps $\tau^{U}$
and $\Lambda^{U}$ we obtain
$2\tau_{1}^{U}(\varrho_{\mathrm{sep}})\geq 2\tau_{2}^{U}(\varrho_{\mathrm{sep}})\geq\Lambda_{2}^{U}(\varrho_{\mathrm{sep}})$,
which leads to another map
$\tilde{\Lambda}^{U}(X)=(\Tr\,X)\mathbbm{1}_{d}+\tau^{U}(X)-X$. It
may be easily shown that $\tilde{\Lambda}^{U}$ is positive but not
CP. For both maps considered in this example $\Lambda_{2}$ from
Eq. (\ref{decomposition}) is an identity map. Thus, due to
previous remarks the resulting inequalities can be applied to
arbitrary states.

{\it Example 4.} Here we discuss two families of indecomposable maps
analyzed in \cite{Tanahasi,Osaka,Ha1} and in \cite{Ha2,Choi,genChoi}.
The first class of maps \cite{Tanahasi,Osaka,Ha1} acts on a $d\times d$ matrix $X$ as
\begin{equation}\label{phidk}
\varphi_{d,k}(X)=(d-k)\epsilon(X)+\sum_{i=1}^{k}\epsilon(S^{i}X
S^{i\dagger})-X,
\end{equation}
where $S$ denotes a shift operator, i.e., $S\ket{i}=\ket{i+1}$,
where addition is understood mod $d$ for $i=1,\ldots,d$. Here
$\epsilon$ stands for a CP map defined as
$\epsilon(X)=\sum_{i=1}^{d}\bra{i}X\ket{i}\proj{i}$. For
$k=1,\ldots,d-2$ the maps $\varphi_{d,k}$ were shown to be
indecomposable \cite{Ha1}, for $k=0$ one has a CP map, $k=d-1$
gives a reduction map $R$, and $\varphi_{3,1}$ is the Choi map
\cite{Choi}. Form (\ref{decomposition}) of $\varphi_{d,k}$ follows
directly from its definition (\ref{phidk}) and reads
$\varphi_{d,k}=\varphi_{d,k}^{(1)}-I$, where $\varphi_{d,k}^{(1)}$
is CP for $k=0,\ldots,d-1$ due to complete positivity of
$\epsilon$ map.

As the second class we consider the maps investigated in Ref.
\cite{Ha2}. Let $a$ and $c_{1},\ldots, c_{d}$ be positive real
numbers. Then the maps for $d\times d$ matrices are defined as
$\Theta\equiv\Theta[a;c_{1},\ldots,c_{d}]=\Theta_{1}[a;c_{1},\ldots,c_{d}]-I$,
where $\Theta_{1}\equiv\Theta_{1}[a;c_{1},\ldots,c_{d}]$ is
defined as
\begin{equation}
\Theta_{1}(X)=a
\epsilon(X)+\mathrm{diag}[c_{d},c_{1},\ldots,c_{d-1}]\epsilon(SX
S^{\dagger}).
\end{equation}
For instance, in this notation $\Theta[2;1,1,1]$ is the Choi map
and its further generalization studied in \cite{genChoi} is
$\Theta[a;c_{1},c_{2},c_{3}]$. The map $\Theta$ is positive iff
$(c_{1}\ldots c_{d})^{1/d}\geq d-a$ and $a\geq d-1$, and
indecomposable if additionally $a<d$. Moreover, $\Theta_{1}$ is
completely positive whenever taken with nonnegative parameters
$a,c_{1},\ldots,c_{n}$ \cite{footnote}.

As presented above both maps $\varphi_{d,k}$ and $\Theta$ have an
identity map as the second CP map in Eq. (\ref{decomposition}).
Therefore, again the resulting inequalities can serve as the
separability criterion for all bipartite states.

{\it Comparison.-} Below we compare the inequalities of type (i)
and (ii) from the Theorem, derived from some of the positive maps
considered in the previous section. We use two classes of states,
namely the SO$(3)$-invariant bipartite states (see, e.g., Refs.
\cite{B1,rot} and references therein for some results concerning
separability properties of this class) and the class of states
considered in Ref. \cite{H-states}.

Every SO$(3)$-invariant bipartite state is a convex combination of
projections $P_J$ onto common eigenspaces of the square of the
total angular momentum and its $z$ component. Here, $J$ takes the
values $|j_A-j_B|,\ldots,j_A+j_B$, where $j_{A(B)}$ denotes the
angular momentum of the subsystem $A(B)$. We focus on the case
when $j_A=j_B=3/2$. Then an arbitrary state may be written as
$\varrho(p,q,r)=p P_{0}+q P_{1}+r P_{2}+s P_{3}$,
where $p,q,r,s=1-p-q-r\in [0,1]$. Entanglement for this states was
characterized in Ref. \cite{B1}, and can be fully described by the
transposition map $T$ and the map $\Lambda^{V}$ ($V$ is an
antisymmetric and anti-diagonal unitary matrix with elements $\pm
1$). Moreover, the states have maximally mixed subsystems and
commute with their partial time reversal
$[I\ot\tau^{V}](\varrho)$. Therefore, they are appropriate to
apply the inequalities derived from the reflection map $\tau^{V}$.

\begin{figure}[]
\includegraphics[width=0.43\textwidth]{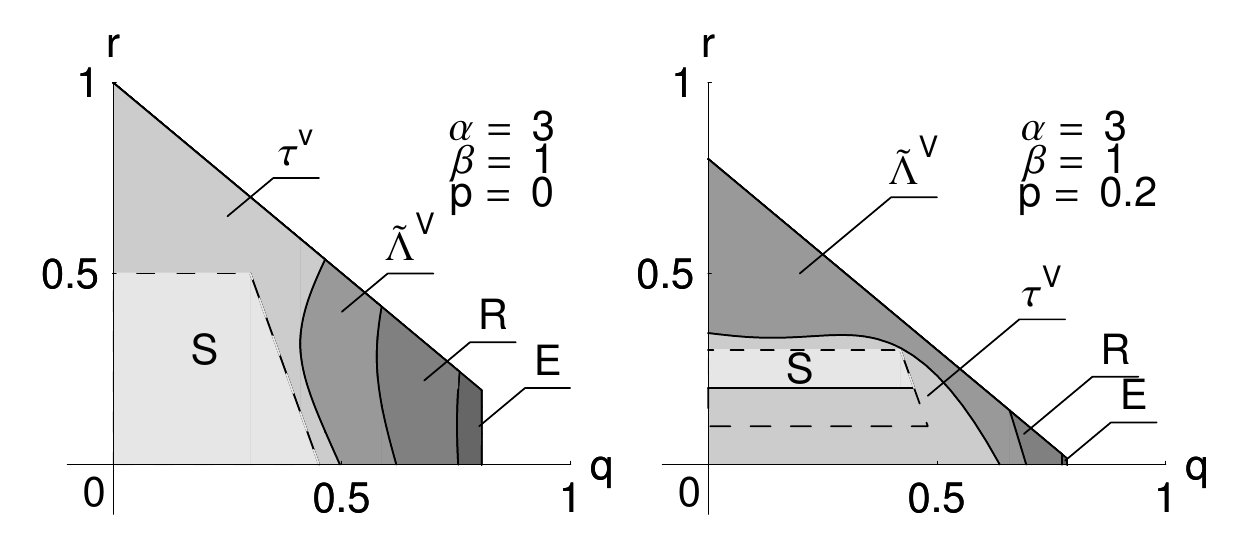}\\
\includegraphics[width=0.43\textwidth]{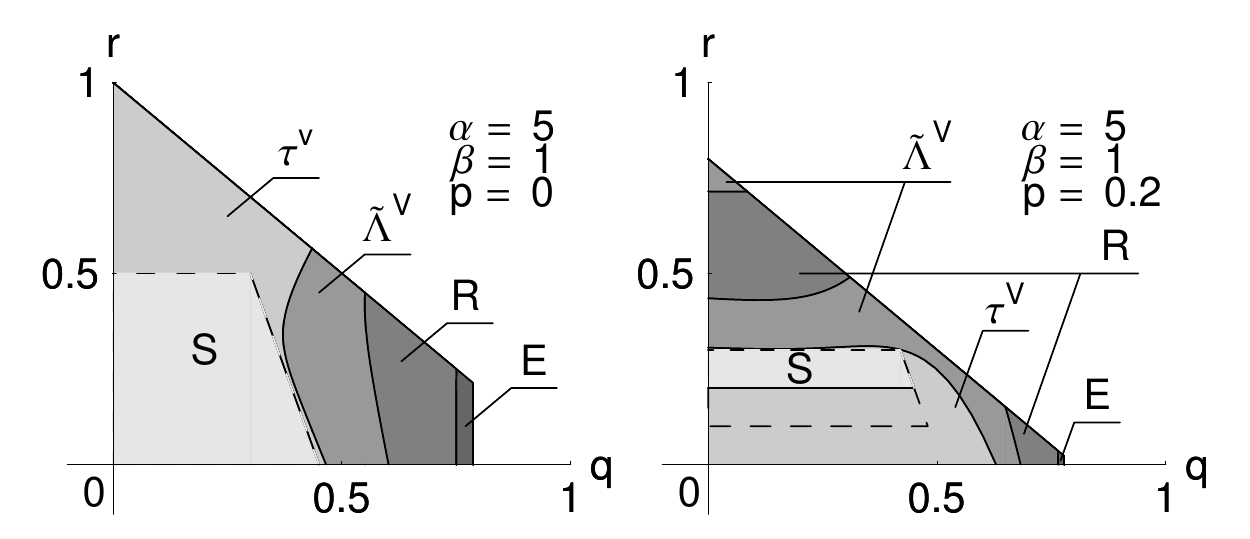}\\
\caption{Comparison of inequalities derived from reduction map
$R$, reflection map $\tau^{V}$, and map $\tilde{\Lambda}^{V}$ from
Example 3 with the classic entropic inequalities. The areas for
which respective inequalities are fulfilled are marked with $R$,
$\tau^{V}$, $\tilde{\Lambda}^{V}$, and $E$. The set of separable
states is the light gray area marked with $S$ and the dashed line
is the border of the set of PPT states. Note that $S\subseteq
\tau^{V} \subseteq \tilde{\Lambda}^V \subseteq R \subseteq E$. The
dimensionless parameters $p$, $q$, and $r$ characterize the
states, whereas $\alpha$ and $\beta$ characterize the
inequalities. The figures are made for different values of
parameters $p$, $\alpha$, $\beta$ given in the
pictures.}\label{fig1}
\end{figure}

In Fig. \ref{fig1} we compare the $(\alpha,\beta)$-inequalities of
type (ii) derived from the reflection map $\tau^{V}$, map
$\tilde{\Lambda}^{V}$ from Example 3, and reduction map $R$ for
integer $\alpha\geq 1,\beta=1$, with the corresponding entropic
inequalities (the $(\alpha=1,\beta\geq 1)$-inequalities derived
from $R$, with the same power $\alpha+\beta$). Note that here one
may look at $\alpha+\beta$ as the number of copies necessary to
perform a collective measurement. As can be seen the region
detected by each inequality (the region where the inequality is
violated) becomes larger with the growth of parameter $\alpha$.

\begin{figure}[]
\includegraphics[width=0.49\textwidth]{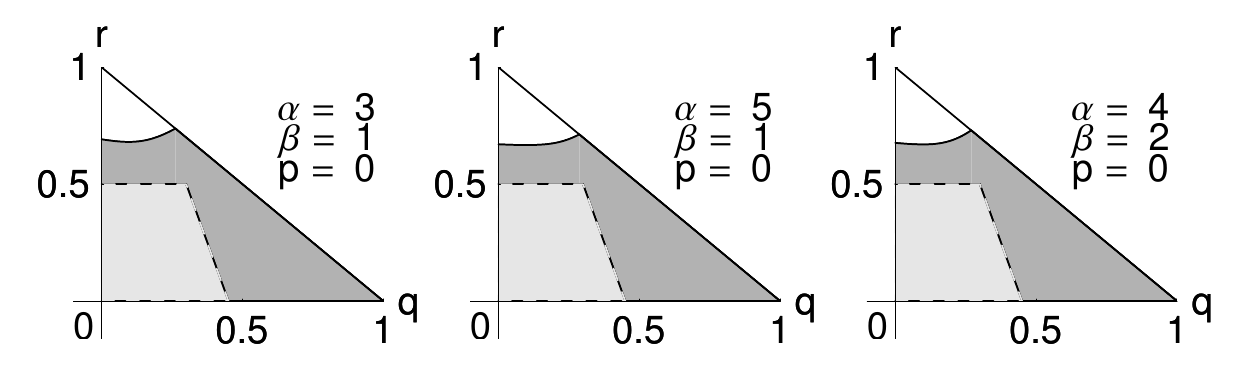}\\
\includegraphics[width=0.48\textwidth]{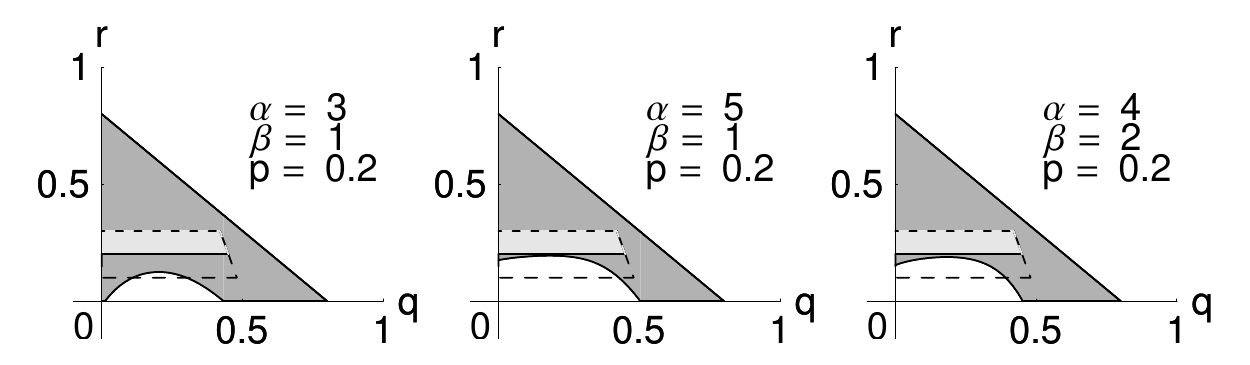}\\
\caption{The region of separable states is marked with light gray,
the dashed line is the border of the set of PPT states. In the
dark gray area the respective inequalities (of type (i) and (ii))
derived from the Breuer map are fulfilled. The figures are made
for few values of parameters $\alpha$, $\beta$ as given in the
pictures, and state parameter $p=0.2$.}\label{fig2}
\end{figure}

In Fig. \ref{fig2} we present the effectiveness of the
inequalities of type (i) and (ii) derived from the map
$\Lambda^{V}$ for a few values of parameters $\alpha$ and $\beta$.
One sees that the detected region becomes larger with the growth
of parameter $\alpha$. Moreover, for the same power $\alpha+\beta$
the inequality with larger $\alpha$ detects entanglement more
effectively. It is interesting that even for small values of
parameter $\alpha+\beta$ $(\alpha=3,\beta=1)$ some PPT
entanglement is detected.

The second class of states we consider was introduced in Ref.
\cite{H-states} and is parameterized as follows
\begin{equation}\label{stanH}
\sigma_{\gamma}=(1/7)\left[\proj{\psi_{+}^{(3)}}+\gamma\sigma_{+}+(5-\gamma)\sigma_{-}\right],
\end{equation}
where $\gamma\in [2,5]$,
$\ket{\psi_{+}^{(3)}}=(1/\sqrt{3})\sum_{i=0}^{2}\ket{ii}$,
$\sigma_{\pm}$ are defined as
$\sigma_{+}=(1/3)\left(\proj{01}+\proj{12}+\proj{20}\right)$
and $\sigma_{-}=\mathscr{V}\sigma_{+}\mathscr{V}^{\dagger}$ with
$\mathscr{V}$ being a bipartite swap operator. The states are
entangled for $\gamma\in (3,5]$ and PPT for $\gamma\in [2,4]$.
They can be detected by the Choi map described in Example 4 either
as $\varphi_{3,1}$, or $\Theta(2;1,1,1)$.
The resulting $(\alpha,\beta)$-inequalities of type (ii)
effectively detect both NPT and PPT entanglement in this class of
states. The range of parameter $\gamma$, for which the
inequalities are violated, versus the values of $\alpha,\beta$ are
given in Table \ref{tab1}.
\begin{table}
\begin{tabular}{@{}c|c|r@{, }l @{}}
  \hline  \hline
  $\qquad \alpha \qquad$ & $\qquad \beta \qquad$ &  \multicolumn{2}{c}{$\qquad\mathrm{Range}$ $\mathrm{of}$ $\gamma\qquad$}\\
  \hline
  6 & 1 &   \multicolumn{2}{c}{--}\\
  7 & 1 & $\qquad(3.191$ & 3.942) \\
  10 & 1 & (3.016 & 4.683) \\
  13 & 1 & (3.002 & 5.0] \\
  $\infty$ & 1 & (3.0& 5.0]\\
  \hline  \hline
\end{tabular}\caption{Range of parameter $\gamma$ of the states given by Eq. (\ref{stanH}),
for which the $(\alpha,\beta)$-inequality of type (ii) derived
from the map $\varphi_{3,1}$ is violated, versus parameters
$\alpha$ and $\beta$.}\label{tab1}
\end{table}

{\it Conclusions.-} In the paper we have provided a method for
derivation of scalar separability criteria from positive but not
CP maps. As a particular example, when one uses the reduction map
\cite{reduction,CAG} and puts $\alpha=1$, the construction leads
to the common scalar separability criteria, known as entropic
inequalities \cite{HHHEntr,reduction,VollbrechtWolf}. Therefore,
it may be treated as their generalization. However, for the
studied positive maps and states the $(\alpha,\beta)$-inequalities
with $\alpha>1$ are much stronger than the entropic ones.
Moreover, as shown in the case of SO$(3)$ invariant states and
states considered in Ref. \cite{H-states}, the inequalities
arising from indecomposable maps detect a large share of bound
entangled states for sufficiently large $\alpha$ (see Fig. 2 and
Table \ref{tab1}), and in the limit $\alpha\to \infty$, the whole
set.

In an attempt to explain how the obtained inequalities work we
considered the limit $\alpha\to\infty$ of the inequalities of type
(ii) with fixed $\beta=1$. This leads to a condition
$\Tr[I\ot\Lambda](\varrho)P_{\mathrm{max}} \geq 0$, where
$P_{\mathrm{max}}$ is such a projector $P$ that corresponds to a
maximum eigenvalue of $\varrho$ and for which
$\Tr[I\ot\Lambda](\varrho)P\neq 0$. Thus, whenever
$P_{\mathrm{max}}$ is an entangled state detected by the conjugate
map $[I\ot\Lambda^{\dagger}]$ the above condition can be treated
as a mean value of a kind of state-dependent entanglement witness
$\mathcal{W}=[I\ot\Lambda^{\dagger}](P_{\mathrm{max}})$. It seems
that at least for some classes of states such witness can detect
entanglement.

An important advantage of the $(\alpha,\beta)$-inequalities of
type (i) and (ii) with integer parameters is that they naturally
lead to a many-copy entanglement witnesses (see e.g.
\cite{ASH,collective}). Moreover, the given examples show that a
relatively small number of copies, meaning the sum $\alpha+\beta$,
is required to effectively detect entanglement. For instance, it
is possible to detect bound entanglement with four copies of a
state (see Fig. \ref{fig2}).

Another point that should be stressed here is that many positive
maps give rise to inequalities that can be applied without the
assumption of commutativity. It would be interesting to
investigate which maps (except for these discussed in Examples
1-4) lead to inequalities applicable to all states.
For instance, the maps considered in Example 4 belong to a general
class analyzed in Ref. \cite{Kossakowski}, namely maps that act as
$\varphi(\ket{i}\bra{i})=\sum_{j}a_{ij}\proj{j}$ and
$\varphi(\ket{i}\bra{j})=-\ket{i}\bra{j}$ for $i\neq j$. They can
be written as $\varphi=\varphi_{1}-I$, where
$\varphi_{1}(\proj{i})=\sum_{j}(a_{ij}+\delta_{ij})\proj{j}$. One
easily finds that $\varphi_{1}$ is completely positive iff
$a_{ij}+\delta_{ij}\geq 0$ for all $i,j$ \cite{footnote}.
However, exact conditions that should be imposed on $a_{ij}$ to
obtain a positive but not CP map $\varphi$, while $\varphi_{1}$
remains CP, are not specified. This, in the context of the
proposed inequalities, would be an interesting subject for further
research.

{\it Acknowledgements.-} Discussions with P. Horodecki and W. A.
Majewski are gratefully acknowledged. This work was supported by
the Polish Ministry of Science and Higher Education under the
grant No. 1 P03B 095 29 and EU Integrated Project SCALA. R. A.
gratefully acknowledges the support of Foundation for Polish
Science.


\begin{thebibliography}{0}
\bibitem{indecomp}The map is called indecomposable iff it cannot be written
as a sum of a completely positive and a completely co-positive
map.
\bibitem{footnote}As proven in Ref. \cite{reduction} a linear map $\Lambda$
is CP if and only if $[I\ot\Lambda](\proj{\psi_{+}^{(d)}})\geq 0$,
where $\ket{\psi_{+}^{(d)}}=(1/\sqrt{d})\sum_{i=0}^{d-1}\ket{ii}$.
\end{thebibliography}
\end{document}